\documentclass[aps,prb,twocolumn,nopacs,amsmath,amssymb]{revtex4}
\usepackage{dcolumn}
\usepackage{bm}
\usepackage{subfigure}
\usepackage{graphicx}
\usepackage{multirow}
\usepackage{hyperref}
\usepackage{lipsum}
\usepackage{hhline}
\usepackage[usenames,dvipsnames,svgnames]{xcolor}

\begin{document}
	
	\makeatletter
	\newcommand*{\balancecolsandclearpage}{%
		\close@column@grid
		\clearpage
		\twocolumngrid
	}
	\makeatother

	\title{Disorder robustness and protection of Majorana bound states in ferromagnetic chains on conventional superconductors}
	\author{Oladunjoye A. Awoga}
	\author{Kristofer Bj{\"o}rnson}
	\author{Annica M. Black-Schaffer}
	\affiliation{Department of Physics and Astronomy, Uppsala University, Box 516, S-751 20 Uppsala, Sweden}
	\begin{abstract}
		Majorana bound states (MBS) are well-established in the clean limit in chains of ferromagnetically aligned impurities deposited on conventional superconductors with finite spin-orbit coupling. Here we show that these MBS are very robust against disorder. By performing self-consistent calculations we find that the MBS are protected as long as the surrounding superconductor show no large signs of inhomogeneity. We also find that longer chains offer more stability against disorder for the MBS, albeit the minigap decreases, as do increasing strengths of spin-orbit coupling and superconductivity.
	\end{abstract}
	\maketitle
	
\section{Introduction}
	The last few years have seen an immense interest in topological superconductors hosting Majorana bound states (MBS) at their boundaries. \cite{qi2011topological, leijnse2012introduction, aliceaReview} A MBS is the condensed matter quasiparticle equivalent of the Majorana fermion theorized to exist in particle physics\cite{wilczek2009majorana, elliott2015RMP} and is, as such, its own antiparticle. MBS also obey non-Abelian statistics in two dimensions (2D), opening the door for fault-tolerant quantum computation. \cite{kitaev2001unpaired, stern2013topological, sarma2015majorana} 
	
	The progress has especially been fueled by the possibility to engineer systems with MBS using relatively common ingredients: superconductivity, spin-orbit coupling, and magnetism. Multiple different avenues are being explored, such as hybrid structures between conventional superconductors (SCs) and spin-orbit coupled semiconducting nanowires\cite{lutchyn2010, oreg2010helical, alicea2010majorana, mourik2012, stanescu2013majorana} or topological insulators \cite{fu2008superconducting} with an applied magnetic field, or by depositing chains of ferromagnetically aligned impurities on a SC with spin-orbit coupling. \cite{choy2011, nadj2013proposal, nadj2014observation} However, as with all materials, disorder effects are inevitable and an outstanding question is thus the stability of the MBS in the presence of disorder.
	Conventional $s$-wave SCs are well-known to be very stable to disorder, as originally established by Anderson, \cite{ANDERSON195926} but the MBS require an effective spinless $p$-wave superconducting state, \cite{readgreen2000paired, leijnse2012introduction, aliceaReview} which could then result in disorder sensitivity. At the same time, a MBS is topologically protected, enforced in the clean system by a finite spectral minigap to other quasiparticle excitations. 
	
	In nanowire-SC hybrid systems it has been emphasized that the mean-free path in the nanowire needs to exceed the proximity-induced coherence length for disorder stability. \cite{brouwer2011topological, sau2012experimental, degottardi2013majorana, cai2013topological} At the same time, disorder in the SC has just recently been predicted to be detrimental if the coupling between the nanowire and SC is beyond the weak-coupling limit,  \cite{DasDisConc, ColeDasnanowire2016} which is desired in order to increase topological protection. \cite{chang2015hardEpitaxial, albrecht2016exponential} 
Ferromagnetic (FM) impurity chains on a SC is in a sense in the extreme coupling limit, with the magnetic moments directly imprinted on the SC. Thus, extrapolating previous result would imply that MBS in FM chains are extremely sensitive to disorder.
				
	In this work we show that the MBS at the end points of FM impurity chains deposited on a SC are actually exceedingly robust against disorder. This is remarkable, especially considering the dramatic disorder vulnerability of the Yu-Shiba-Rusinov (YSR) states \cite{yu1965, shiba1968classical,rusinov1969theory} created by single magnetic impurity. \cite{DasDisConc} 
	More specifically, by performing self-consistent calculations also taking into account the effect of disorder on the superconducting state, we find that the MBS exist and are stable as long as the surrounding (conventional $s$-wave) SC  shows only moderate signs of inhomogeneity. Thus simply measuring the local density of states (LDOS) in the surrounding SC gives a very good prediction of the MBS disorder stability.
	By increasing the chain length we find that the disorder stability of the MBS is even further increased, although the minigap becomes somewhat more disorder sensitive. We also observe that, similar to related systems,\cite{sau2012experimental} increased spin-orbit coupling and superconductivity also enhance disorder robustness.
	
\section{Method}
	We here focus on chains of magnetic impurities forming a FM chain deposited on a 2D $s$-wave SC with Rashba spin-orbit interaction, see insert in Fig.~\ref{fig:delta006}(a). A topological non-trivial phase with MBS at the FM chain end points are well established theoretically \cite{choy2011, nadj2013proposal, braunecker2013interplay, klinovaja2013topological,vazifeh2013selforganized, pientka2013helicalShiba,bjornson2015spin}  and credible experimental MBS signatures have recently been reported for FM Fe chains deposited on Pb, which is an $s$-wave SC with significant surface spin-orbit coupling. \cite{nadj2014observation, RubyFranke2015, Pawlak2016, feldman2016highresolution}
	The full Hamiltonian can be written as a combination of the SC and FM parts as $\hat{H} = \hat{H}_{\textrm{SC}} + \hat{H}_{\textrm{FM}}$, which are effectively modeled on a square lattice as\cite{nadj2013proposal,pientka2013helicalShiba,bjornson2016majorana}
	%
	\begin{align}\label{eqn:HamiltonianSC}
		\hat{H}_{\textrm{SC}} &= -t\sum_{\langle i,j \rangle,\sigma} c_{i\sigma}^{\dag}c_{j\sigma} -\sum_{ i,\sigma} (\mu-\delta\mu_i) c_{i\sigma}^{\dag}c_{i\sigma}  \nonumber \\
		&-\alpha\sum_{i\mathbf{r}} (e^{i\theta_\mathbf{r}}c_{i+\mathbf{r}\downarrow}^{\dag}c_{i\uparrow} +\textrm{H.c.}) + \sum_ i\Delta_i(c_{i\uparrow}^{\dag} c_{i\downarrow}^{\dag}+ \textrm{H.c.}), \nonumber \\
		\hat{H}_{\textrm{FM}} &= -V_Z\sum_{ p,\sigma,\sigma'} (\sigma_z)_{\sigma\sigma'}c_{p\sigma}^{\dag}c_{p\sigma'},
	\end{align}
	where the operator $c_{i\sigma}^{\dag}$ creates a particle with spin $\sigma$ at site $i$. Here $t$ is the nearest neighbor hopping, $\mu$ is the overall chemical potential, and $\alpha$ is the strength of the Rashba spin-orbit coupling, which has directional dependences given by the polar coordinates $\theta_\mathbf{r}$ of the nearest neighbor bond vectors $\mathbf{r}$. 
	Conventional $s$-wave superconductivity is represented by an on-site order parameter $\Delta_i$. It is common to simply assume a constant order parameter  $\Delta_i = \Delta$. However, this approach fails to take into account the influence of both magnetic impurities and disorder on the superconducting state. We therefore also perform self-consistent calculations, where a site-dependent $\Delta_i$ is allowed. In a self-consistent framework only a constant pair potential $V_{\textrm{sc}}$, representing the fixed strength of the pair-pair interactions, is assumed. The Hamiltonian Eq.~\eqref{eqn:HamiltonianSC} is then solved by diagonalization within the Bogoliubov-de-Gennes framework \cite{deGennes} and the order parameter is reiteratively calculated using $\Delta_i=-V_{\textrm{sc}}\langle c_{i\downarrow}c_{i\uparrow}\rangle$ until convergence \cite{BlackSchaffer2008self, bjornson2013vortex, bjornson2016majorana}. This is particularly important here as it guarantees that disorder effects are appropriately accounted for in the SC. Moreover, it also captures the suppression, and even sign reversal, of the order parameter near the magnetic impurities. \cite{Salkola1997spectral, Flatte1997local,meng2015superconducting, bjornson2016piphase} We note that this suppression is very laterally limited, and thus using a 2D sheet for the SC is well justified. 
	For the FM chain we assume large and well-ordered magnetic moments, ignoring quantum fluctuations. The magnetic impurities will then to a good approximation only give rise to an effective Zeeman term $V_Z$ on each impurity site $p$ in the SC.
	We construct FM chains of different lengths by forming linear chains of magnetic impurities, while always ensuring that the surrounding SC is large enough such that the outer boundaries do not influence our results. We use $t = 1$ and the lattice distance as natural units and set $\mu=-4t$ to create a normal state with a finite density of states. While we do not explicitly use parameter values for Fe on Pb, our results cover both this and other systems. 
	
	After achieving a stable topological superconducting phase with clear MBS at the chain end points in the clean system, we introduce disorder and study its effects. We focus on generic potential disorder in the SC in the form of a site-dependent chemical potential distributed randomly, $\delta\mu_i \in[-w,w]$. Such Anderson-type disorder is generated by charge inhomogeneities or puddle formation and has the benefit of preserving the effective chemical potential. We have also compared with results for finite concentrations of point scatterers with constant strength $\delta\mu_i = w$, but find no significant differences, see Supplementary Material (SM) for details.\cite{SM} Assuming that the magnetic impurity chain forms a stable FM state, disorder in the SC is the only significant disorder source in this system. 
	For each set of parameters we use $128$ different disorder configurations and average over all configurations. To quantify the effects of disorder on the MBS, we mainly focus on the two lowest (positive) energy levels. The lowest energy level is the MBS energy $E_M$, while the second lowest level is the minigap $E_G$, protecting the MBS from quasiparticle poisoning. In what follows $E_{M/G}$ are disorder-averaged values unless otherwise stated, while $E_G^0$ is the clean system minigap.
	We first focus on a constant fixed order parameter and study the effect of disorder as we vary chain length and all tunable parameters: $\Delta$, $V_Z$, or $\alpha$. We then turn to the more accurate self-consistent calculations. 
	
	\section{Non-Self-consistent Results}
	\subsection{Chain length}
	For the non-self-consistent calculations we first consider FM chains of different lengths, while keeping all other parameters fixed such that the clean system is well within the topological phase. 
	\begin{figure}[htb]
		\centering
		{
			\includegraphics[]{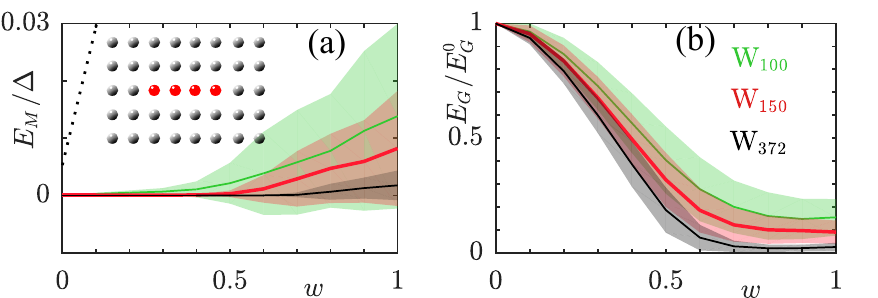}
		}
		\caption{(Color online.) MBS (a) and minigap (b) energies as a function of disorder strength $w$ for different chain lengths with the YSR energy (dotted). Shaded regions show one standard deviation. Here  $\alpha=0.28$, $\Delta=0.06$, and $V_z=0.88$. Inset shows a  schematic of a $W_4$ chain, with SC sites with (without) magnetic impurities in red (black), however, there are many more surrounding SC sites to avoid influence from the outer edges.
		}
		\label{fig:delta006}
	\end{figure}
	%
	In Fig.~\ref{fig:delta006} we show $E_{M}$ and $E_G$ for three different FM chains: W$_{100}$, W$_{150}$, and W$_{372}$, with subscripts denoting chain length, i.e.~the number of magnetic impurities in the chain in units of the SC lattice constant. These are also compared to the YSR state for a single magnetic impurity (dotted). As seen, there is a huge increased disorder stability of the MBS compared to the YSR state. In fact, the YSR state continues approximately linearly as a function of $w$ even beyond the plot, while all chains show only very moderate deviation from zero energy even at the strongest disorder levels, where $w =1$ is already $25 \%$ of the band width and $\sim 17\Delta$. Furthermore, longer chains offer notably more protection of the MBS against disorder than shorter chains. This is not only seen in the disorder-averaged $E_M$, but also in the decrease of the spread, given by one standard deviation (shaded regions), with increasing chain lengths. The reason for this behavior is the larger hybridization between the two end point MBS in shorter chains, which is enhanced with disorder, see Fig.~\ref{fig:ProbabibilityDensity}. As seen, the MBS always hybridize at high enough disorder strengths $w$, but the hybridization is much more pronounced for the shorter chain in Fig.~\ref{fig:ProbabibilityDensity}(a) compared to the longer chain in Fig.~\ref{fig:ProbabibilityDensity}(b).
	%
	\begin{figure}[htb]
		\centering
		{
			\includegraphics[width=8.5cm]{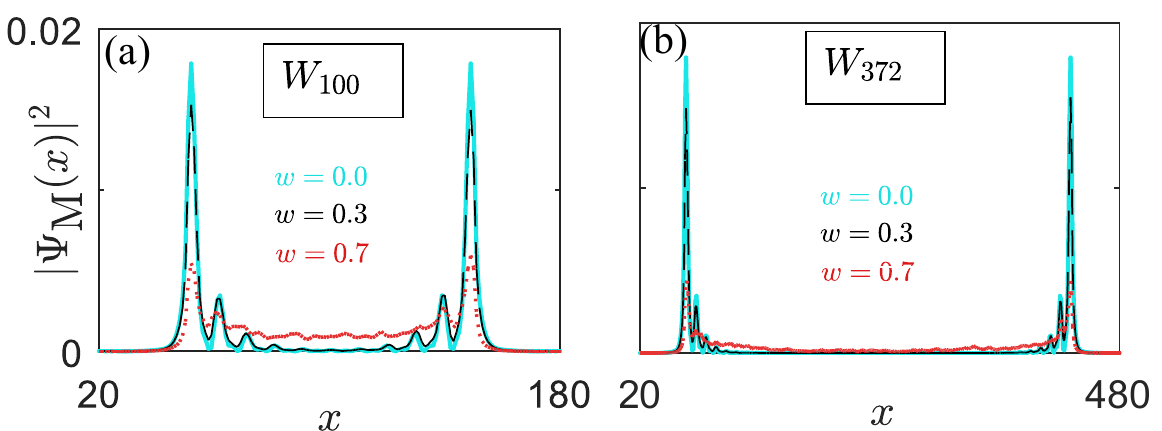}
			}
		\caption{(Color online.) Disorder-averaged probability density of the MBS eigenstate for $\alpha=0.28$, $\Delta=0.06$, and $V_z=0.88$ for several disorder strengths $w$ and with chain lengths W$_{100}$ (a) and W$_{372}$ (b).
		}
		\label{fig:ProbabibilityDensity}
	\end{figure}

	Taken together, the results shows that MBS are very protected against disorder in contrast to individual YSR states that are highly susceptible to disorder. The protection is further improved by increasing the chain length, which has also been found to be true for  nanowires proximity-coupled to a SC in the absence of disorder. \cite{MishmashAlicea2016}
	
	While the $E_M$ level is very protected against disorder, the minigap is more affected, at least at strong disorder. As seen in Fig.~\ref{fig:delta006}(b), $E_G$ is reduced with increasing $w$, an effect enhanced in longer chains. This can be understood by noting that the number of impurity-induced subgap states increases linearly with chain length, and therefore there are more states that are prone to disorder widening in longer chains. \footnote{The increased number of subgap states also influences the clean minigap $E_G^0$, with longer chains having smaller minigaps, although for long enough chains $E_G^0$ saturates at a lower bound. \cite{MishmashAlicea2016}}
	The decrease in $E_G$ with disorder is eventually leveling off. A minimum value is reached (around $w\geq 0.9$ for  W$_{100}$) when the first and second energy levels start to interact. However, an actual level crossing is always avoided due to finite spin orbit coupling, \cite{bjornson2016piphase} which causes a flat, or upturn, behavior of $E_G$ with increasing disorder. 
	There is thus a trade-off in disorder stability between short and long chains. A long chain clearly offers more protection for the MBS, while its minigap is reduced faster with increased disorder. Still, the disorder spread is more limited for longer chains, making the behavior of single samples more reproducible.
		This chain length dependence does not qualitatively depend on the size of the order parameter $\Delta$ (see SM\cite{SM}).
	
	%
	\subsection{Spin-orbit coupling}
\begin{figure}[htb]
		\centering
		{
			\includegraphics[width=8.5cm]{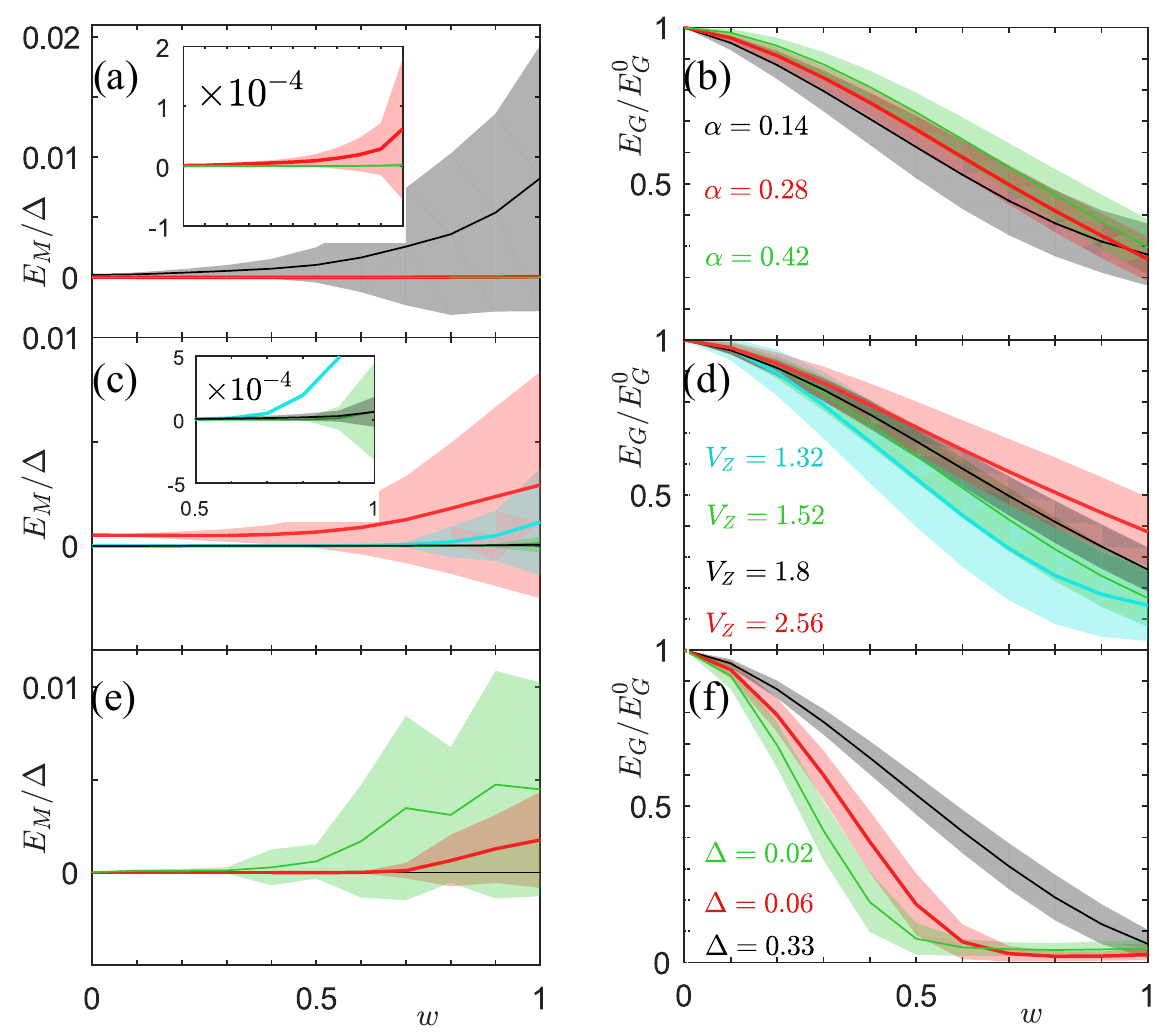}
		}
		\caption{(Color online.) MBS (a,c,e) and minigap (b,d,e) energies as a function of disorder strength $w$. Varying  $\alpha$ for $\Delta=0.33$, $V_Z = 1.52$ with a W$_{100}$ chain (top), varying $V_Z$ for $\alpha=0.28$, $\Delta = 0.33$ with a W$_{100}$ chain (middle), and varying $\Delta$ for $\alpha=0.28$ at optimal $V_Z$ (see main text), with a W$_{372}$ chain (bottom). Shaded regions show one standard deviation. Insets show a zoom-in at small energies.}
		\label{fig:ParamVar}
	\end{figure}

	Next we investigate the behavior when varying all the parameters, starting with the spin-orbit coupling $\alpha$. In this case both $E_{M}$ and $E_G$ show disorder robustness with increasing $\alpha$, see Figs.~\ref{fig:ParamVar}(a,b). This is a consequence of the topological phase in the clean limit  being more protected for larger $\alpha$: $E_G^0$ is proportional to $\alpha$, \cite{sau2012experimental,cayao2015sns, MishmashAlicea2016, bjornson2016majorana} while the MBS localization length decreases with increasing $\alpha$ making the MBS level splitting smaller. \cite{MishmashAlicea2016, sarma2012splitting} The reduced level splitting results in the MBS being  less sensitive to disorder for larger spin-orbit coupling.
	%
	%
		\subsection{Zeeman energy}
		We also tune the strength of the magnetic impurities $V_Z$, while staying within the topological regime in the clean case. Figure~\ref{fig:ParamVar}(c) shows a non-monotonic disorder sensitivity of $E_M$ with varying $V_Z$. This effect can be attributed to the behavior in the clean limit. 
		For the highest $V_Z$ reported there is significant hybridization between the two end point MBS in the clean case, which leads to reduced disorder stability. On the other hand, for the lowest $V_Z$, the system is relatively close to the topological phase transition and the MBS are thus not very localized, which also increases disorder sensitivity. The best disorder stability is therefore achieved for moderate $V_Z$, where the system is well within the topological phase but the MBS still only suffer minimal hybridization, see also SM. \cite{SM}
		The minigap behavior, Fig.~\ref{fig:ParamVar}(d), is more simple with more disorder stability for increasing $V_Z$, such that when entering deeper into the topological phase the minigap is more robust against disorder. This is despite the fact that $E_G^0$ is generally reduced with $V_Z$ beyond the topological phase transition region \cite{MishmashAlicea2016, SM} and shows that effective disorder stability is not completely determined by the clean limit. 
		
	\subsection{Superconducting order parameter}
	Finally we study disorder robustness for different $\Delta$. However, changes in $\Delta$ result in the system moving within the topological phase diagram, \cite{bjornson2015spin} as with the $V_Z$ changes in Fig.~\ref{fig:ParamVar}(c,d). To isolate disorder effects we therefore choose for each $\Delta$ the $V_Z$ value yielding optimal topological stability in the clean limit.
	As seen in Figs.~\ref{fig:ParamVar}(e,f), larger $\Delta$ clearly offers more protection against disorder, both for the MBS and minigap. 
	For systems with weak superconductivity it is thus more important to have long chains as longer chains prevent the end point MBS from hybridizing, even with the increasing SC coherence length for decreasing $\Delta$. Note that although the MBS localization length have been found to be notably renormalized away from the SC coherence length, it is still highly dependent on it. \cite{peng2015strong} 
	Also notable is that the disorder strength for the weakest $\Delta$ in Figs.~~\ref{fig:ParamVar}(e,f) is in fact a significantly higher multiple of $\Delta$ than for the stronger SCs, e.g.~$w=0.6$ is actually $w=30\Delta$ for $\Delta =0.02$ but only $w=1.8\Delta$ for $\Delta=0.33$. Thus, the MBS are surprisingly robust against disorder even for weak SCs, as long as the disorder is not multiple magnitudes larger than the bulk superconducting gap. 
	
	\section{Self-consistent results}
	So far we have assumed a constant superconducting order parameter $\Delta_i=\Delta$, but this is a crude approximation, especially when disorder is present. We therefore also carry out self-consistent calculations to capture how $\Delta_i$ is affected by disorder and how that in turn influences the MBS and the minigap.
	Due to drastically increasing computational complexity we are now limited to only study shorter chains: W$_{18}$  and W$_{28}$, while setting $V_{\textrm{sc}}$ such that $\Delta =0.3$ in the clean bulk, away from magnetic impurities and outer boundaries. However, we find that other $V_\textrm{sc}$ give qualitatively similar results.
	
\begin{figure}[htb]
	\centering
		{
		\includegraphics[width=7cm]{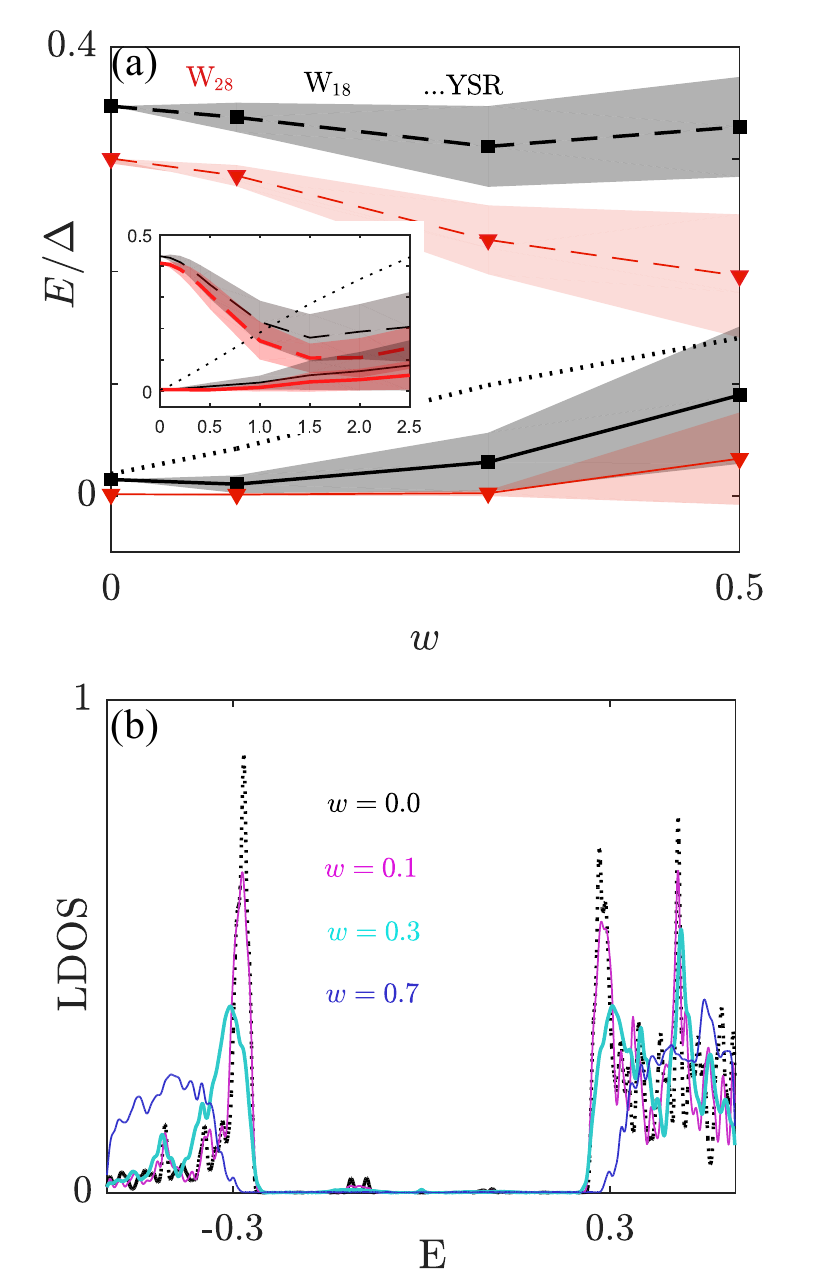}
		\caption[mini]{(Color online.) Self-consistent results for MBS (solid lines), minigap (dashed lines) and YSR (dotted line) as a function of disorder strength $w$ (a) and disorder-averaged LDOS at a single bulk site for a $W_{28}$ chain (b). Shaded regions in (a) show one standard deviation. Inset: Non-self-consistent results for the chain. Here $\alpha =0.28$, $V_z=1.0$, $V_{\textrm{sc}}=4.34$, resulting in $\Delta_i=0.3$ in the clean bulk. LDOS is plotted with a Gaussian smoothing and enhanced by a factor of $10^3$.}
	\label{fig:Self}
	}
	\end{figure}

	Figure~\ref{fig:Self}(a) shows how the MBS and minigap energies change with increased disorder when $\Delta_i$ is determined self-consistently compared to the constant $\Delta$ case. The overall behavior is similar to the earlier non-self-consistent results: the MBS are more protected against disorder in longer chains, while the minigap then becomes more disorder sensitive. The YSR state from a single impurity is likewise much more affected by disorder than any MBS.
		%
	%
	\begin{figure*}[htb]
		\centering
		{
			\includegraphics[]{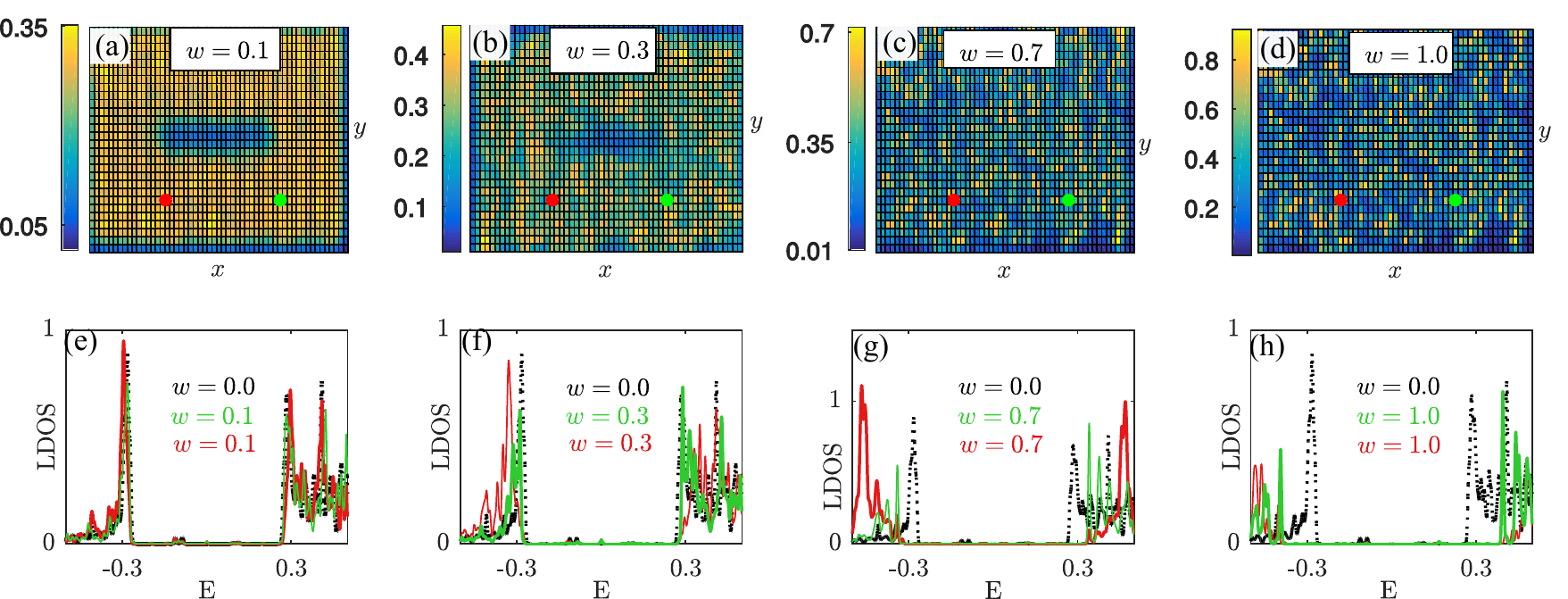}
			\caption[Order parameter]{(Color online.) Self-consistent results for a W$_{28}$ chain in a single disorder realization showing $\Delta_i$ (top) and LDOS at two different bulk sites (bottom). Sites used for the LDOS in (e-h) are indicated with red/green filled circles in (a-d). Here $\alpha =0.28$, $V_z=1.0$, $V_{\textrm{sc}}=4.34$, resulting in $\Delta_i=0.3$ in the clean bulk. LDOS is plotted with a Gaussian smoothing and enhanced by a factor of $10^3$.}
			\label{fig:SelfOP}
		}
	\end{figure*}
	However, beyond overall similarities, self-consistency has a notable effect. Figure~\ref{fig:Self}(a) shows that the MBS in a $W_{18}$ chain start to significantly couple to other states already around $w \approx 0.5$, seen  clearly in the upturn of $E_G$, which signals level repulsion. The same situation does not appear in the non-self-consistent calculation until $w \approx 1.5$, as shown in the insert of Fig.~\ref{fig:Self}(a). Thus including self-consistency seems to significantly reduce disorder stability. However, this is a much too naive conclusion. In fact, we will now show that self-consistent results generate a very viable experimental condition for when MBS are stable in FM chain systems.
	
	In order to understand disorder stability it is necessary to first study the bulk properties of the SC. With finite disorder present the order parameter varies throughout the lattice, but after disorder averaging these fluctuations are notably reduced or even completely washed out (see SM\cite{SM}). Thus conclusions based on disorder-averaged $\Delta_i$ can be misleading. Instead we focus on analyzing single disorder realizations, which is also most relevant experimentally considering that a measurement primarily takes place on a single sample. Figures~\ref{fig:SelfOP}(a-d) show $\Delta_i$ in the whole system for a single disorder realization for multiple different $w$. At low disorder strengths the almost complete suppression of superconductivity around the chain found in clean systems \cite{Balatskyet.alRMP2006,meng2015superconducting,bjornson2016piphase} is clearly visible. Still, $\Delta_i$ is not constant in the bulk even for the smallest disorder used. At intermediate $w = 0.3$ there is a notable non-uniformity in $\Delta_i$, but superconductivity is still strong throughout the system and the suppression of the order parameter at and around the wire is still visible.
However, as $w$ increases further, the distribution of $\Delta_i$ becomes highly nonuniform and there will be regions where $\Delta_i \approx 0$ and other regions consisting of superconducting islands with very large $\Delta_i$. In fact, the position of the chain is not even noticeable in $\Delta_i$ for $w \gtrsim 0.5$. The cross-over between a stable superconducting substrate and that of a heavily disordered SC broken up into multiple superconducting islands is thus around $w = 0.5$. Similar strong variation of $\Delta_i$ with disorder has recently also been reported for uniform (i.e.~not a chain) magnetic coverage.\cite{qin2016disorder}
	
	The strong non-uniformity in $\Delta_i$ leaves also clear fingerprints in the LDOS. This is  illustrated very clearly in Figs.~\ref{fig:SelfOP}(e-g), where we plot the LDOS at the two different bulk sites indicated by red and green filled circles in Figs.~\ref{fig:SelfOP}(a-d). 
	At $w=0.1$ the variations in energy and shape of the superconducting coherence peaks are negligible and at $w=0.3$ there are still clearly identifiable coherence peaks, albeit their energies are slightly shifted between different regions. Thus the also LDOS here signals the stability of the superconducting bulk phase.
	However, with further increased disorder the LDOS starts to vary strongly between different sites and also compared to the clean spectrum. For $w \gtrsim 0.5$ it is clear that the substrate SC is now extremely non-uniform, with both energy and shape of the superconducting coherence peaks varying strongly, producing a highly non-uniform SC with notably different local energy gaps. Thus, the larger disorder strengths used here are clearly much larger than those tolerated according to Anderson's theorem for $s$-wave SCs.
	
	To further demonstrate the disorder effects on the bulk superconducting state, we plot in Fig.~\ref{fig:Self}(b) the disorder-averaged LDOS in the bulk. The averaging procedure not only consolidates the results from many different possible disorder realizations, but is also an effective and experimentally measurable quantity of the average disorder in a single sample.
	As $w$ gradually increases the height of the coherence peaks is suppressed, albeit at first BCS-like coherence peaks are preserved. However, at $w \gtrsim 0.5$ the coherence peaks are replaced with a dome-like structure.
	Note that although there are regions of $\Delta_i\approx 0$ at high disorder, the spectral gap in these regions is still non-vanishing because $\delta\mu_i$ is in these regions very large which means only high energy states can exist there. \cite{ghosal1998role} 
	
	The strong spatial variations of both $\Delta_i$  and LDOS at high disorder are very useful for determining the disorder stability of the MBS for a FM chain. Figures~\ref{fig:Self} and~\ref{fig:SelfOP} show that the MBS are only effectively lost even in shorter chains for $w \gtrsim 0.5$. At the same time, at these disorder strength the bulk SC shows very clear signs of strong disorder. We  thus conclude that MBS are robust as long as the surrounding SC does not show signs of strong disorder. The latter can be determined straightforwardly by measuring the LDOS using scanning tunneling spectroscopy (STS), but also direct observations of local order parameter variations have recently been demonstrated using scanning Josephson spectroscopy. \cite{randeria2016scanning} If the LDOS show clear coherence peaks with non-dispersive energies, then well-protected MBS should be formed at the end points of a FM chain positioned well within the topological phase.
	
	
	\section{Conclusions}
	In summary we have found that MBS formed at the end points of a FM magnetic impurity chain deposited on a SC with spin-orbit coupling are surprisingly protected against disorder, especially considering the notable vulnerability of the single impurity YSR states. Longer chains leave the MBS more protected although the minigap to other quasiparticle excitations then decreases somewhat faster with disorder.
	By performing self-consistent calculations we have shown that the stability of the MBS can be predicted simply by measuring the level of disorder in the SC: as long as the surrounding SC show relative homogeneity in the LDOS, well-protected MBS will be present within the topological phase.
	%
	
	\acknowledgments
	This work was supported by the Swedish Research Council (Vetenskapsr\aa det), the Swedish Foundation for Strategic Research (SSF), the G\"{o}ran Gustafsson Foundation, and the Wallenberg Academy Fellows program through the Knut and Alice Wallenberg Foundation.
	\bibliographystyle{apsrevmy}
	\bibliography{abbrRef,Ref}
	
	\balancecolsandclearpage
	\onecolumngrid
	\begin{center} {\large \bf Supplementary material}\end{center}
	\section*{Supplementary Material}
In this supplementary material we provide additional figures and accompanied discussion to further support the work and conclusions of the main text. We also present results using single-type disorder, i.e.~varying the concentration for fixed impurity strength, instead of the Anderson-type disorder used in the main text. 

%
\subsection{Chain length}
\begin{figure*}[htb]
	{
		\includegraphics[]{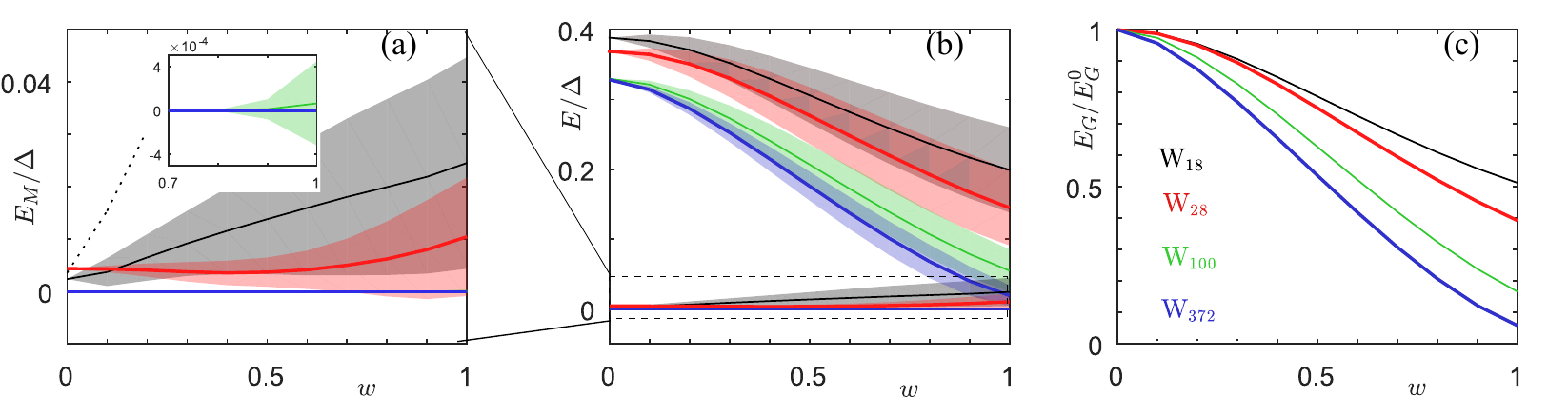}
	}
	\caption{(Color online.) MBS (a) and minigap (b,c) energies as a function of disorder strength $w$ for different chain lengths with the YSR energy (dotted). Shaded regions show one standard deviation. Here $\alpha=0.28$, $\Delta=0.33$, and $V_z=1.52$. The minigap is plotted either normalized to $\Delta$ and directly compared to the MBS energy (b) or to the clean mingap $E_G^0$ (c). Inset shows a zoom-in at small energies. 
	}
	\label{fig:Delta03}
\end{figure*}

In the main text we use a superconducting order parameter $\Delta=0.06$ when studying how different chain lengths influence the disorder stability. 
To further investigate the effect of chain length, we also here also report results for $\Delta=0.33$ using four different chain: W$_{18}$, W$_{28}$, W$_{100}$, and W$_{372}$. As shown in Fig.~\ref{fig:Delta03},the results are qualitatively the same as for the smaller superconducting order parameter. 
%

It is worth explaining why the minigap at zero disorder $E_G^0$ (Fig.~\ref{fig:Delta03}(b)) changes with the length of the chain. Due to the finite length of the chains, there is no sharp transition into the topological phase, but rather a smooth cross-over is observed. As the chain length increases, the first excited stated moves closer and closer to zero on the trivial side of the transition point before opening into $E_G^0$ on the topological side of the transition point (see also Fig.~\ref{fig:EnergyLevels} below). This is the reason behind different $E_G^0$ with varying chain length. For long enough chains, the topological phase transition becomes sharp, and $E_G^0$ is then constant, provided all other parameters are fixed. This is why $E_G^0$ for W$_{100}$ and W$_{372}$ are nearly identical. This is in agreement with an earlier study of clean wires~\cite{MishmashAlicea2016}.

\subsection{Varying parameters}
\begin{figure*}[htb]
	\centering
	{
		\includegraphics[]{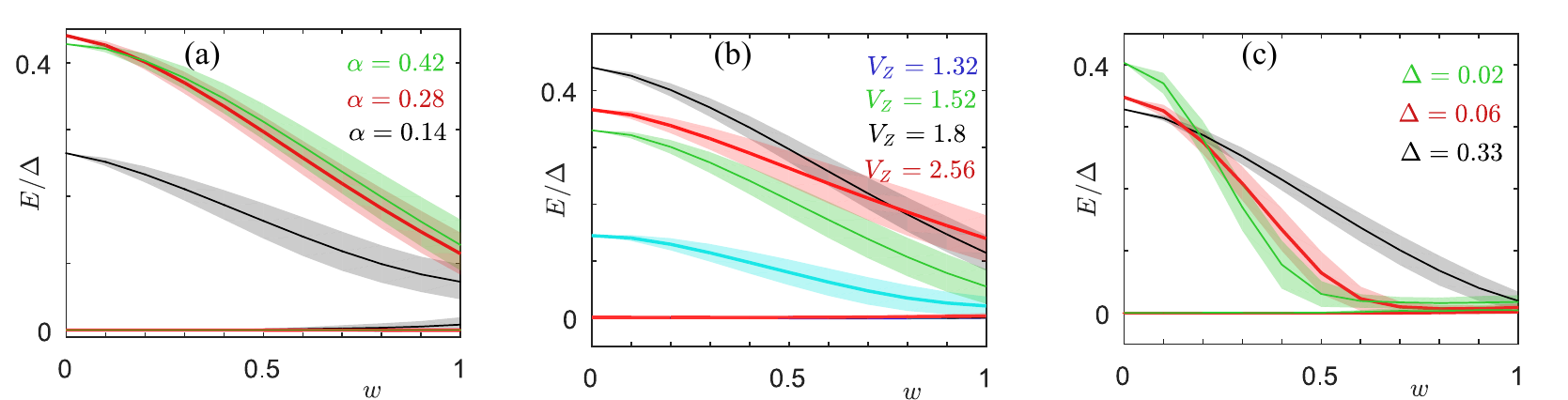}
	}
	\caption{(Color online.) $E_G$ and $E_{M}$ (thin lines) in units of $\Delta$ for the same parameters as in Fig.~3 of the main text.}
	\label{fig:Unnorm}
\end{figure*}

When varying $\alpha$, $V_Z$, and $\Delta$ in the main text we only presented $E_G/E_G^0$, i.e.~we normalized the minigap by its clean value in order to most clearly display the effect of disorder. But in some cases the minigap in the clean system changes when these parameters are varied. In Fig.~\ref{fig:Unnorm} we for completeness display these changes by instead plotting $E_G$ in units of $\Delta$, while also plotting $E_{M}$ for a direct comparison. As seen, the MBS are for almost all disorder strengths well separated from all other states and, while the absolute values of $E_G$ changes, there is no change in the overall disorder behavior. 

\begin{figure*}[htb]
	\centering
	{
		\includegraphics[]{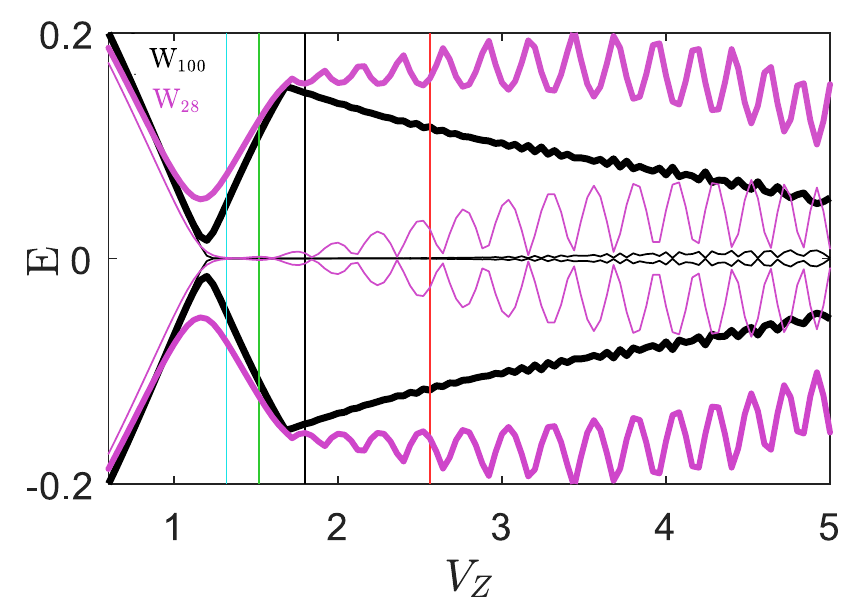}
	}
	\caption[Magnetic chain]{(Color online.) $E_M$ (thin) and $E_G$ (thick) as a function of $V_Z$ for two different chain lengths in a clean system. Vertical lines, with corresponding colors, indicate the values of $V_Z$ used in Figs.~2(c,d) in the main text and also used in Fig.~\ref{fig:ProbVz}. Here $\alpha=0.28$ and $\Delta=0.33$.}
	\label{fig:EnergyLevels}
\end{figure*}

\begin{figure*}[htb]
	\centering
	{
		\includegraphics[]{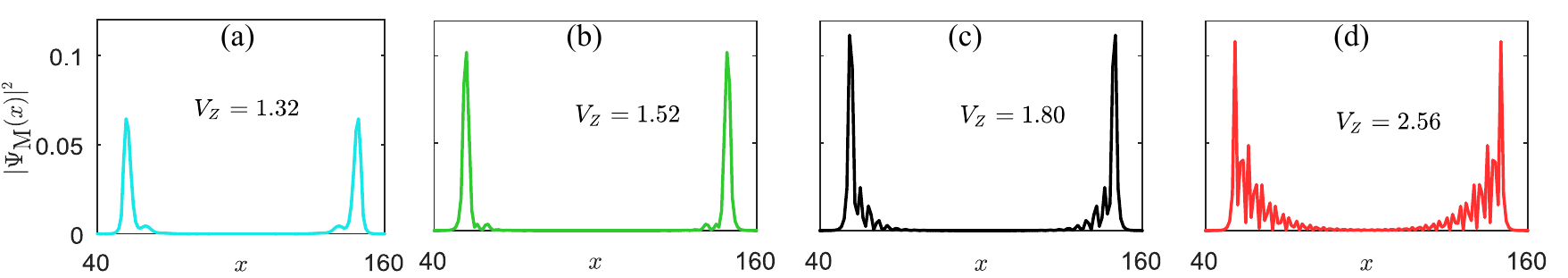}
	}
	\caption[Magnetic chain]{(Color online.) Probability density of the MBS for several values of $V_z$ in a clean system.  Here  $\alpha=0.28$ and $\Delta=0.33$ with a W$_{100}$ chain. The different colors correspond to the vertical lines in Fig.~\ref{fig:EnergyLevels}.}
	\label{fig:ProbVz}
\end{figure*}

In the main text we also briefly discuss how the MBS and minigap behave as a function of the magnetic term $V_Z$ in the clean system. This is illustrated in detail in Figs.~\ref{fig:EnergyLevels} and \ref{fig:ProbVz} for each value of $V_z$ used in Figs.~2 (c,d) of the main text and also in Fig.~\ref{fig:Unnorm}(b). 
Figure \ref{fig:EnergyLevels} clearly show how the topological phase transition is a smooth cross-over for short chains and only for longer chains does the bulk energy gap approximately close at the topological phase transition. It also shows how $E_G$ generally decreases with chain length in the non-trivial phase, both in the phase transition region and beyond. This  explains the non-monotonic behavior found for $E_G^0$ as a function of $V_Z$ in Fig.~\ref{fig:Unnorm}(b).
In terms of the MBS behavior, there is clearly an intermediate regime of $V_Z$ where the MBS are closest to zero energy. At small $V_Z$, just beyond the topological phase transition, the finite energy splitting is due to the localization being weaker in the finite sized phase transition region, as illustrated in Fig.~\ref{fig:ProbVz}(a). At large $V_Z$ the two chain end point MBS start to hybridize with each other as evident both in Fig.~\ref{fig:ProbVz}(d) and in the notable oscillations of their energy levels in Fig.~\ref{fig:EnergyLevels}. These effects reduce disorder stability both in the small and large $V_Z$ limits as discussed in the main text, and there is thus an intermediate $V_Z$ regime with best disorder stability. 

\subsection{Self-consistent calculations: Disorder-averaged order parameters}
Self-consistent results were reported for chains $W_{18}$ and $W_{28}$ with a bulk order parameter converging to $\Delta = 0.3$ in Figs.~ 4 and 5 of the main text.

\begin{figure*}[htb]
	\centering
	{
		\includegraphics[]{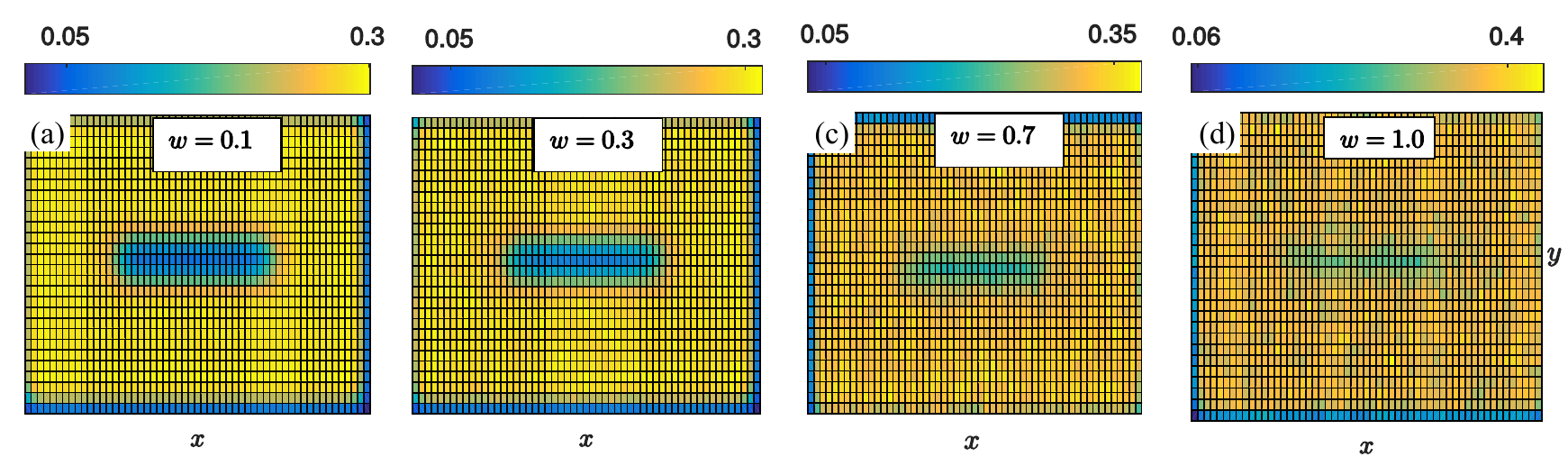}
	}
	\caption[Magnetic chain]{(Color online.) Disorder-averaged self-consistently determined $\Delta_i$ with a W$_{28}$ chain for multiple values of disorder strength $w$. Here  $\alpha=0.28$, $V_Z=1.0$, and $V_{sc}=4.34$, resulting in $\Delta_i=0.3$ in the clean bulk.}
	\label{fig:OPavg}
\end{figure*}

Considering that any experimental measurement primarily takes place on a single sample, studying single disorder realizations is most relevant. However, for completeness we also show the disorder-averaged order parameter for a range of disorder strengths in Fig.~\ref{fig:OPavg}. All parameters here are the same as that of Fig.~5 in the main text where a single disorder realization is studied.
For $w \leq 0.3$ there is not much difference compare to the clean case.
However, for higher $w$, even the disorder-averaged order parameters in Figs.~\ref{fig:OPavg}(c,d) start to change from site to site, albeit the variation is clearly significantly washed out (due to the disorder averaging) compared to a single disorder realization. Thus even for disorder-averaged $\Delta_i$ (and LDOS) there are clear signs of heavy disorder effects in the superconductor. This verifies that different samples will show the same qualitative behavior.
\\
\\
\\
\subsection{Single Impurity Disorder}
In the main text we focused on Anderson disorder, i.e.~random site-dependent fluctuations of the chemical potential such that $\delta \mu_i \in [-w,w]$ on each site. Such fluctuations are naturally occurring due to charge inhomogeneities or puddle formation. They also have the added benefit of keeping the effective chemical potential constant in the whole sample, which leads to better control of the topological phase. For comparison we here also report results for a single type of impurity, such that the chemical potential is changed by a fixed amount $\delta \mu_i = w$, but only on randomly selected sites with a fixed disorder concentration $n$.
This type of disorder has been considered for topological superconducting nanowires \cite{DasDisConc} but not, to the best of our knowledge, for magnetic impurity chains deposited on a superconducting surface. 

\begin{figure*}[htb]
	{
		\includegraphics[]{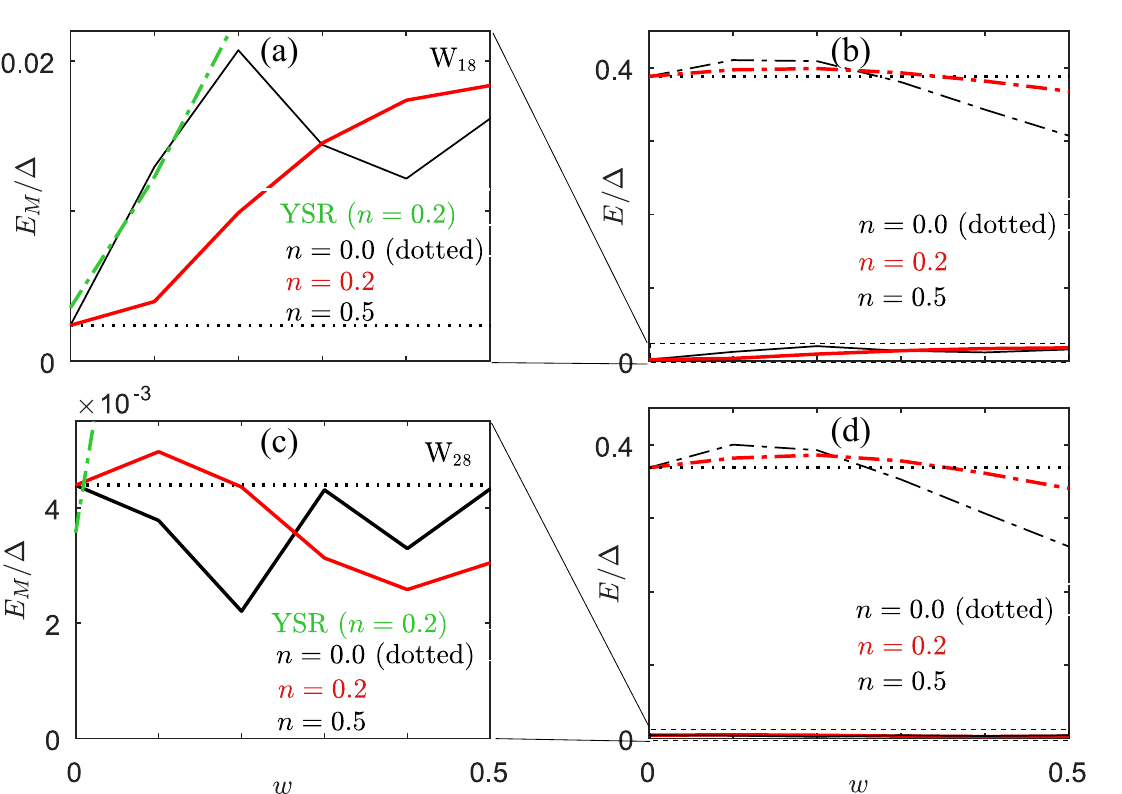}
		\caption[Disorder with concentration]{(Color online.) MBS (a,c) and minigap (b,d)  energies as a function of disorder strength $w$ for W$_{18}$ (top) and  W$_{28}$ (bottom) chains and for varying disorder concentration $n$.  (a,c) are zoom-ins of the dashed low energy regions in (b,d), respectively. Here $\alpha=0.28$, $\Delta=0.33$, and $V_z=1.52$.}
		\label{fig:DisConc}
	}
\end{figure*}

Figure~\ref{fig:DisConc} show the results for two chains W$_{18}$ and W$_{28}$ using the same parameters as in Fig.~\ref{fig:Delta03} for a direct comparison. As seen, the YSR state is very unstable even with only 20\% disorder at low disorder strengths $w$. The MBS at the FM chain end points are however much more stable, and especially so in longer chains. 
For the minigap we actually see a small increase at weak disorder strengths, and even more so for larger disorder concentrations. This is due to the effective chemical potential increasing with increased disorder concentration, which leads the system into a more stable part of the topological phase. This effect is only visible at low disorder strengths, such that the disorder still does not cause a suppression of $E_G$. The increased overall chemical potential and the accompanied increase in $E_G$ also enhance the stability of the MBS, especially in longer chains where the MBS are already quite stable against the disorder. 
Overall, disorder effects using single impurities are quite similar to that of Anderson disorder: The MBS are remarkably robust against disorder, with longer FM chains displaying even more protection in terms of the MBS energy levels, while the minigap eventually decreases for strong disorder and more so for longer chains. However, single impurities cause additional effects by changing the effective chemical potential in the sample, which can mask the true effects of disorder.
Finally, self-consistent calculations do not give any additional information to what is already presented in the main text for Anderson disorder.

\end{document}